\documentclass[12pt]{article}
\usepackage{amsfonts}
\usepackage{amssymb}
\begin{document}
\date{}
\title{\textbf{Remarks on the Generalized Chaplygin Gas}}
\author{{Rabin Banerjee}$^1$\thanks{E-mail: rabin@bose.res.in}, \ {Subir Ghosh}$^2$\thanks{E-mail: sghosh@isical.ac.in} \ and \ {Shailesh Kulkarni}$^1$\thanks{E-mail: shailesh@bose.res.in}\\
$^1$\textit{S.~N.~Bose National Centre for Basic Sciences,}\\
\textit{JD Block, Sector III, Salt Lake, Kolkata-700098, India}\\
$^2$\textit{Physics and Applied Mathematics Unit, Indian Statistical Institute,}\\
\textit{203 B.T Road, Kolkata 700108, India}}

\maketitle
\begin{quotation}
\noindent \normalsize
We have developed an action formulation for the Generalized Chaplygin Gas
 (GCG). The most general form for the nonrelativistic GCG action is derived
consistent with the equation of state. We have also discussed a relativistic
formulation for GCG by providing a detailed analysis of the Poincare algebra.
\end{quotation}
\section{Introduction}
The recent observation of accelerated expansion of the universe,
concluded \cite{obs} from the study of luminosity of type Ia
distant supernova, has put Cosmology in the center stage. Our
inability to explain the origin of this expansion has led to the
naming of this  phenomenon as "Dark Energy" effect. The coinage 
obviously matches the
other fuzzy area in Cosmology, {\it{i.e.}} the existence of
"Dark Matter".There exist several plausible models at hand that can explain the
astronomical data \cite{obs}. The traditional one - vacuum energy
or non-zero cosmological constant - fits well with the
observational data. Unfortunately it is plagued with serious
conceptual difficulties: smallness of the value of the
cosmological constant in comparison with Planck mass scale and the
coincidence problem, (that questions the reason for the near
equality between energy densities of Dark Energy and dust-like
matter in the present epoch), to name a few. The latter is
circumvented by introducing scalar field (or Quintessence) models
\cite{quint} inducing a dynamical vacuum energy, but only at the
expense of fine tuning the scalar potential parameters. An
alternative dynamical model \cite{kam} for Dark Energy, featuring
Chaplygin Gas \cite{chap} or its generalization \cite{gcg} - the
Generalized Chaplygin Gas (GCG) - has created a lot of interest in
recent times. Conventional analysis of the model \cite{ch1} has
attributed its matter density profile the flexibility to smoothly
interpolate between a dust dominated era (at early times) to the
Cosmological constant dominated era (at present times). A further
generalization \cite{bilic} to inhomogeneous GCG allows one to
address the issue of Dark Matter  as well. The GCG model has
passed several experimental tests of various nature, such as high
precision Cosmic Microwave Background Radiation data \cite{exp1},
supernova data \cite{exp2} and gravitational lensing \cite{exp3}.
Naive analysis \cite{teg} seemed to suggest  a disturbing
phenomenon in the GCG model: possible existence of unphysical
oscillations or even an exponential blow-up in the matter power
spectrum at present. However, this problem has been solved in
\cite{sol} by taking in to account the interaction between Dark
Matter, Dark Energy and phantom-type Dark Energy \cite{Xin}. (For a detailed
exposition of these issues, see \cite{orf}.)

The above mentioned ideal fluid system  was introduced long ago by
Chaplygin \cite{chap} as an effective model in computing the
lifting force on a wing of an airplane. It obeys an  exotic
equation of state,
\begin{equation}
P=-\frac{B}{\rho} \label{16}
\end{equation}
where $P(x)$ and $\rho (x)$ denote pressure and density
respectively and $B$  a constant parameter.  However, the interest in Chaplygin gas model
 actually goes beyond Cosmology (see \cite{jac} for a review, oriented towards the High Energy Physics community).
 It has a deep connection with the $D$-branes in a higher dimensional Nambu-Goto Formulation in light-cone
 parameterization \cite{bord}. It is also unique in admitting a supersymmetric generalization \cite{jac1} for a fluid. The dynamical role of Chaplygin gas
in cosmology has been shown in \cite{bg}.  The above discussion clearly underlines the relevance and topicality of GCG models in Cosmology and High Energy Physics.
 
 In the present work, we follow the same theme and focus our attention on the GCG models, where the generalization amounts to
 postulating the Chaplygin equation of state as,
\begin{equation}
P = -\frac{B}{\rho^{\alpha}};~~B>0;~~0<\alpha <1  \label{1.1}
\end{equation}
where the standard Chaplygin pressure equation (\ref{16}) is recovered for
$\alpha =1$. In the nonrelativistic regime, we have constructed the most 
general action for GCG consistent with (\ref{1.1}). This construction has been done in two versions. In one case,
the action involves the density $\rho$ and the velocity potential $\theta$.
Elimination of $\rho$ is possible leading to the second version involving only 
$\theta$. This may be interpreted as a Born-Infeld type action in the nonrelativistic limit

    Next, the construction of relativistic action for GCG is discussed.
Here a Born-Infeld action involving only $\theta$ is proposed which
has the correct nonrelativistic limit. Also for $\alpha =1$, it reproduces
the standard Born-Infeld action for the Chaplygin gas. We may mention
that our action is different from the one given in the literature \cite{gcg}.
Since the density plays an important role it becomes worthwhile to write
the relativistic action for GCG involving both $\rho$ and $\theta$ analogous
to the usual $\alpha =1$ case \cite{jac}. However here we are faced with certain problems.
Our suggested form for GCG action has the correct nonrelativistic and
$\alpha =1$ limits. However for $\alpha \ne 1$ it is relativistic only for large $\rho$. This is found by an explicit check of the Poincare algebra.

The paper is organized as follows: In Section {\bf{2}}  we provide
a brief review of the non-relativistic and  relativistic action
formulations of the normal Chaplygin gas ($\alpha =1$). This helps
us to fix the notation and charts the course of our subsequent
analysis in the paper. Section  {\bf{3}} comprises an analogous study
for the GCG ($\alpha \ne 1$) and introduces new expressions for the
nonrelativistic GCG action. Section {\bf{4}} is devoted to the construction and
subsequent  analysis of the relativistic GCG. In Section {\bf{5}} we provide our conclusion and
propose avenues for future study.

\section{Normal ($\alpha =1$) Chaplygin gas : a brief review}
Before concentrating on the Chaplygin gas, let us discuss some basic notions of fluid dynamics, in the Lagrangian
(action) formulation \cite{jac}.

   We start with the non-relativistic scenario. The equations of motion, governing an ideal fluid in  arbitrary space dimensions,  are given by
\begin{eqnarray}
\partial_t \rho(t, {\bf x}) + \nabla \cdot (\rho(t,{\bf x}){\bf v}(t,{\bf x})) &=& 0 \label{0.1},\\
\partial_t {\bf v}(t,{\bf x}) +  {\bf v}(t,{\bf x})\cdot \nabla {\bf v}(t,{\bf x}) &=& {\bf f}(t,{\bf x}),\label{0.2}
\end{eqnarray}
where  $\rho(t,{\bf x})$ and ${\bf v}(t,{\bf x})$ are the  matter
density  and velocity     fields respectively. The first identity
reflects
     the matter conservation and the second is the Euler equation of
     motion. We consider the motion of fluid to be  isentropic. Hence the force $f$ is
     derived from a $\rho $-dependent potential $V(\rho)$,
\begin{equation}
{\bf f} = -{\bf \nabla}V'(\rho).\label{0.3}
\end{equation}
Here prime denotes derivative with respect to $\rho$. Note that $V'(\rho)$ is the enthalpy. The all important
relation connecting $V$ with the pressure $P$  is given by,
\begin{equation}
P(\rho) = \rho V'(\rho) - V(\rho). \label{0.4}
\end{equation}
In the case of irrotational fluid we can further restrict the
velocity to be a gradient of some scalar field $\theta ({\bf x})$,
\begin{equation}
{\bf v} = {\nabla} \theta.\label{0.5}
\end{equation}
The (non-relativistic) Hamiltonian for irrotational motion is just
the sum of kinetic and potential energy,
\begin{equation}
H = \int dr (\frac{1}{2}\rho (\partial_i \theta)^2 + V(\rho)).
\label{0.6}
\end{equation}
  Now, the first order form of the Lagrangian $L$, corresponding to (\ref{0.6})
 is given by,
\begin{equation}
L=\int dr (\theta \dot \rho -\frac{1}{2}\rho (\partial_i \theta)^2
- V(\rho)). \label{0.6.1}
\end{equation}
From the symplectic structure it is clear that $\rho$ and $\theta$ are conjugate variables,
 satisfying the the canonical Poisson bracket,
\begin{equation}
\{\theta({\bf x}), \rho({\bf y})\} = \delta({\bf x}- {\bf y}). \label{0.9.1}
\end{equation}
The nature of the potential function $V(\rho )$ will specify the particular fluid model under study. For Chaplygin
gas the potential profile is given by,
\begin{equation}
V(\rho) = \frac{\lambda}{\rho }, \label{0.7}
\end{equation}
where $\lambda$ is the interaction strength.  Using (\ref{0.5}) the Lagrangian for the Chaplygin gas model is
given by
\begin{equation}
L=\int dr (\theta \dot \rho -\frac{1}{2}\rho (\partial_i \theta)^2
- \frac{\lambda}{\rho}), \label{0.8}
\end{equation}
 and the Bernoulli equation follows:
\begin{equation}
\dot \theta + \frac{1}{2}(\partial_i \theta)^2 = \frac{\lambda}{\rho ^2}.\label{0.9}
\end{equation}

It is possible to  eliminate  $\rho$ from the Lagrangian to obtain a non-relativistic Born-Infeld like structure
in  $\theta$,
\begin{equation}
L(\theta) = - 2\sqrt{\lambda}\int dr \sqrt{\left(\dot \theta +
\frac{1}{2}(\partial_i \theta)^2\right)}, \label{0.11}
\end{equation}
with the equation of motion,
\begin{equation}
\partial_t\left(\dot \theta + \frac{1}{2}(\partial_i \theta)^2\right)^{-\frac{1}{2}}
+ \partial_i \left[\partial_i \theta \left(\dot \theta + \frac{1}{2}(\partial_i \theta)^2\right)^{-\frac{1}{2}}\right]=0. \label{0.11.1}
\end{equation}

  Now we come to the relativistic generalization of Chaplygin gas\cite{jac}. A Lagrangian has been suggested for the normal $(\alpha =1)$ Chaplygin gas in \cite{jac},
\begin{equation}
L = \int dr (\theta \dot\rho - \sqrt{(\rho^2c^2 + a^2)}\sqrt{c^2 + (\partial_i \theta)^2}), \label{0.12}
\end{equation}
where $a$ is a parameter in the theory. Although (\ref{0.12}) does not have a manifestly relativistic form, its
Poincare invariance has been demonstrated explicitly in \cite{bg} in a Hamiltonian framework (see Section {\bf{4}}
of the present paper as well). On the other hand, one can eliminate $\rho $ once again from (\ref{0.12}) to obtain
the Born-Infeld form,
\begin{equation}
L= -a\int dr \sqrt{c^2 - \partial_{\mu}\theta\partial^{\mu}\theta}
\label{be}
\end{equation}
which is manifestly relativistic.

To get the correct  non-relativistic limit, one has to consider the map \cite{jac},
\begin{equation}
\theta \rightarrow  \theta - tc^2. \label{1.13}
\end{equation}
Under this transformation, the relativistic model (\ref{0.12}) will reduce to the non-relativistic one in
(\ref{0.8}) with the  identification $\lambda \equiv \frac{a^2}{2}$. This concludes our review \cite{jac} of action
formulation of Chaplygin gas.

\section{Nonrelativistic generalized ($\alpha \ne 1$) Chaplygin gas}
In GCG the equation of state (\ref{16}) is replaced by a more flexible one,
 given in (\ref{1.1}). In order to incorporate
this generalization in the action formulation, our starting  point is to find a suitable potential $V(\rho)$
compatible with (\ref{1.1}). To find the general solution for $V(\rho)$
 we start from an ansatz,  
\begin{equation}
 V(\rho) = (\frac{B}{\alpha +1}) \frac{1}{\rho^{\alpha}} + u(\rho)
\end{equation}
where $u(\rho)$ is such that $V(\rho)$ satisfies
\begin{equation}
-\frac{B}{\rho^{\alpha}} = \rho \frac{dV(\rho)}{d\rho} - V(\rho).
\end{equation}
This follows from (\ref{1.1}) and the enthalpy relation (\ref{0.4}). This implies $u(\rho)$ must satisfy
\begin{equation}
\rho \frac{du}{d\rho} - u = 0.
\end{equation}
The solution for above equation is
\begin{equation}
u(\rho) = I \rho. 
\end{equation}
where $I$ is an integration constant.

    Hence the most general form of the potential $V(\rho)$  is
\begin{equation}
V(\rho) = (\frac{B}{\alpha +1})\frac{1}{\rho^{\alpha}} + I\rho.\label{1.3}
\end{equation}

    For the irrotational fluid we can write the Hamiltonian (\ref{0.6}), with  $V(\rho)$ as given in
    (\ref{1.3})
\begin{equation}
H = \int dr (\frac{1}{2}\rho (\partial_i \theta)^2
+\frac{B}{(\alpha +1)\rho^{\alpha}} + I\rho), \label{1.4}
\end{equation}
Now the first order form of the Lagrangian follows from (\ref{1.4}),
\begin{equation}
L^{\alpha} = \int dr (\theta \dot \rho -\frac{1}{2}\rho
(\partial_i \theta)^2 - \frac{B}{(\alpha +1)\rho^{\alpha}} - I\rho),
\label{1.5}
\end{equation}
where the superscript $\alpha$ on $L$ reveals the fact that we are
dealing with GCG. 

  Variation of $\rho$ yields the Bernoulli equation,
\begin{equation}
\dot \theta + \frac{1}{2}(\partial_i \theta)^2 =
\frac{B\alpha}{(\alpha+1)\rho^{\alpha +1}} - I. \label{1.6}
\end{equation}

     To obtain the $\rho$ independent Lagrangian for GCG, one can  use the Bernoulli equation to reexpress   $\rho$ in terms of  $\theta$,
\begin{equation}
\rho = \left(\frac{\alpha B}{\alpha+1}(\dot \theta
+\frac{1}{2}(\partial_i\theta)^2 + I )^{-1}
\right)^{\frac{1}{\alpha+1}}. \label{1.7.1}
\end{equation}
It is very convenient to rewrite the Lagrangian given in
(\ref{1.5}) in the following form
\begin{equation}
L^{\alpha} = -\int dr \left( \dot \theta \rho +
\frac{\rho}{2}(\partial_i \theta)^2 + \frac{B}{(\alpha
+1)\rho^{\alpha}} + I\rho\right). \label{1.8}
\end{equation}
In the above equation we have omitted  total derivative terms. Substituting
 $\rho$ from (\ref{1.7.1}) in (\ref{1.8}) we find, 
\begin{equation}
L^{\alpha}(\theta) = -
\left(\frac{\alpha}{\alpha +1}\right)^{\frac{\alpha}{\alpha+1}}
B^{\frac{1}{\alpha+1}}\int dr \sqrt{\left(\dot\theta +
\frac{1}{2}(\partial_i
\theta)^2 + I\right)^{\frac{2\alpha}{\alpha+1}}}. \label{1.10}
\end{equation}
This is the most general form of GCG Lagrangian and is a central result of
this paper. It is the Born-Infeld version of nonrelativistic GCG.

    The  equation of motion for $\theta $ turns out to be,
\begin{equation}
\partial_t\left(\dot \theta + \frac{1}{2}(\partial_i \theta)^2\right)^{\frac{-1}{\alpha+1}}
+ \partial_i \left[\partial_i \theta \left(\dot \theta
+ \frac{1}{2}(\partial_i \theta)^2 + I\right)^{\frac{-1}{\alpha+1}}\right]=0. \label{1.11}
\end{equation}
A definite simplification occurs by setting $I=0$. Then the Lagrangians
 (\ref{1.5}) and (\ref{1.10}) reduce to  
\begin{equation}
L^{\alpha} = \int dr (\theta \dot \rho -\frac{1}{2}\rho
(\partial_i \theta)^2 - \frac{B}{(\alpha +1)\rho^{\alpha}}),
\label{1.11.1}
\end{equation}

\begin{equation}
L^{\alpha}(\theta) = -
\left(\frac{\alpha}{\alpha +1}\right)^{\frac{\alpha}{\alpha+1}}
B^{\frac{1}{\alpha+1}}\int dr \sqrt{\left(\dot\theta +
\frac{1}{2}(\partial_i
\theta)^2 \right)^{\frac{2\alpha}{\alpha+1}}}. \label{1.11.2}
\end{equation}
Putting $\alpha =1$ in (\ref{1.11.1}) and (\ref{1.11.2}) reproduces 
the expressions for the usual Chaplygin gas \cite{jac}.

\section{Relativistic generalized Chaplygin gas}
Now we turn to the relativistic form of GCG. Any relativistic version of GCG 
must satisfy two conditions: it should have the correct nonrelativistic limit
 (\ref{1.5}) or (\ref{1.10}), secondly, for $\alpha =1$ it should reduce to (\ref{0.8}) or (\ref{0.11}).
 
To begin with we suggest a manifestly Poincare invariant model for GCG, given by\begin{equation}
L^{\alpha} = -(a')^{\frac{1}{\alpha+1}} \int dr \sqrt{(c^2 - \partial_{\mu}\theta\partial^{\mu}\theta)^{\frac{2\alpha}{\alpha+1}}}\label{alphabe} 
\end{equation}
In the nonrelativistic  limit it agrees with (\ref{1.11.2}).To show this we exploit (\ref{1.13}) and use the fact that  $\partial_{\mu}\theta\partial^{\mu}\theta = \frac{\dot \theta^2}{c^2} -\partial_i\theta^2$, to simplify the above Lagrangian, 
\begin{eqnarray}
L^{\alpha} &=& -(a')^{\frac{1}{\alpha+1}}\int dr
\left[-\frac{\dot\theta}{c^2}+2\dot\theta
+ (\partial_i \theta)^2 \right]^{\frac{\alpha}{\alpha+1}}.\nonumber\\
\end{eqnarray}
Now taking the large $c$ limit we get
\begin{eqnarray}
\lim_{c\rightarrow \infty} L^{\alpha} = -
(2)^{\frac{\alpha}{\alpha+1}}(a')^{\frac{1}{\alpha+1}}\int dr \sqrt{\left[\dot \theta+(\frac{\partial_i \theta)^2}{2}\right]^{\frac{2\alpha}{\alpha+1}}}\label{alpha1}.
\end{eqnarray}
By identifying $a' = (\frac{\alpha}{2(\alpha+1)})^{\alpha}B$ we see that (\ref{alpha1}) agrees with  (\ref{1.11.2}).
Also, in the $\alpha =1$ limit our Lagrangian (\ref{alphabe}) reduces to that of usual relativistic Chaplygin gas (\ref{be}). This shows that it is possible to interpret (\ref{alphabe}) as a viable form for the relativistic GCG Lagrangian.

  At this point we should mention that there exists in the literature a Poincare invariant form for GCG \cite{gcg}
\begin{equation}
L_b = - A^{\frac{1}{1+\alpha}} \int dr \left[c^2 -
(\partial_{\mu}\theta
\partial^{\mu}\theta)^{\frac{1+\alpha}{2\alpha}}\right]^{\frac{\alpha}{\alpha+1}}.\label{bertolamiBE}
\end{equation}
Note that, for $\alpha\ne 1$ the above Lagrangian is different from (\ref{alphabe}). However for $\alpha =1$ it agrees with normal Chaplygin gas Lagrangian \cite{jac}. GCG of similar nature \cite{gcg}  coupled to gravity has been considered in \cite{Hess}.  
 
   Now consider the nonrelativistic limit of (\ref{bertolamiBE}). Following 
the same procedure as discussed above we get in this limit 
\begin{equation}
L_b = -A^{\frac{1}{1+\alpha}}(2Y)^{\frac{\alpha}{\alpha+1}}\sqrt{\left[\frac{Z}{2Y} + (\dot\theta+\frac{1}{2}\nabla\theta^2)\right]^{\frac{2\alpha}{1+\alpha}}}.
\label{bertolamiNRBE}
\end{equation}
where
\begin{eqnarray}
Y&=& \frac{1+\alpha}{2\alpha}c^{\frac{1+\alpha}{\alpha}}, \nonumber\\
Z&=& c^2 - c^{\frac{1+\alpha}{\alpha}}. \nonumber
\end{eqnarray}
This Lagrangian is same as that of (\ref{1.10}) provided we identify $I$ with $\frac{Z}{2Y}$. Thus both (\ref{alphabe}) and (\ref{bertolamiBE}) are valid forms for the relativistic GCG whose nonrelativistic limits correspond to different
 parametrizations of the general form for nonrelativistic GCG given in (\ref{1.10}).

 Let us next attempt to construct the relativistic GCG model by including
 the density field $\rho$. Also, since the  density field plays an 
important role in the observational analysis of GCG it is worthwhile to have 
 a relativistic version for GCG involving $\rho$ and the velocity
 potential $\theta$.

   To this end, we consider the following  Lagrangian for relativistic GCG:
\begin{equation}
L^{\alpha} = \int dr \left(\theta \dot\rho - \sqrt{(\rho^2c^2 +
\frac{a^2}{\rho^{\alpha-1}})}\sqrt{c^2 + (\partial_i
\theta )^2}\right), \label{1.12}
\end{equation}
where  $a$ is a constant parameter. To ensure the correct  nonrelativistic limit
we use the same  map as (\ref{1.13}), and  explicitly check that in the  $c\rightarrow \infty$ limit, the nonrelativistic GCG model (\ref{1.11.1}) is reproduced, provided we identify, 
\begin{equation}
a = \sqrt{\frac{2B}{\alpha+1}}. \label{1.16}
\end{equation}
We put $c =1$ and obtain the equations of motion, 
      \begin{equation}
\dot \rho + \partial_i \left( \frac{\sqrt{(\rho^2 +
\frac{a^2}{\rho^{\alpha-1}})}}{\sqrt{1 + (\partial_i \theta)^2}}
\partial_i \theta\right) = 0, \label{1.18}
\end{equation}
\begin{equation}
\dot \theta = -\frac{\sqrt{1 + (\partial_i
\theta)^2}}{\sqrt{(\rho^2 + \frac{a^2}{\rho^{\alpha-1}})}}[\rho
c^2 - (\frac{\alpha -1}{2})\frac{a^2}{\rho^{\alpha}}] .\label{1.19}
\end{equation}
They also have the correct  $\alpha=1$ limit \cite{jac}.

As we have pointed out before, the Lagrangian (\ref{1.12}) has been posited by us in analogy with the relativistic Lagrangian given in (\ref{0.12}) \cite{jac} and we know that (\ref{1.12}) has the correct $\alpha =1$ limit. Since the model (\ref{1.12}) is not manifestly Lorentz invariant, it becomes imperative to check the Poincare algebra. To this end, we follow our previous work \cite{bg} and compute the canonical energy-momentum tensor $T_{\mu\nu}$ (in  the Noether prescription),
\begin{equation}
T_{\mu\nu} =
\frac{\partial\mathcal{L}}{\partial(\partial^{\mu}\psi_{i})}\partial_{\nu}\psi^{i}
- g_{\mu\nu}\mathcal{L}.\label{1.19.1}
\end{equation}
  Using the above definition, the explicit form of the components of $T_{\mu\nu}$ are given by,
\begin{eqnarray}
T_{00}&=& \sqrt{(\rho^2 + \frac{a^2}{\rho^{\alpha-1}})}\sqrt{1 + (\partial_i \theta)^2},\label{1.20}\\
T_{0i}&=& \theta \  \partial_{i} \rho,\label{1.21}\\
T_{i0}&=& -\frac{\sqrt{\rho^2 + \frac{a^{2}}{\rho^{\alpha-1}}}}{\sqrt{1+(\partial_{k}\theta)^2}} (\partial_{i}\theta) \dot \theta = \left(\rho({\bf x}) +
\left(\frac{1-\alpha}{2}\right)\frac{a^2}{\rho^{\alpha}({\bf
x})}\right)\partial_{i}\theta ,\label{1.22}\\
T_{ij}&=& -\frac{\sqrt{\rho^2 +
\frac{a^{2}}{\rho^{\alpha-1}}}}{\sqrt{1+(\partial_{k}\theta)^2}}
(\partial_{i}\theta) (\partial_{j}\theta) -
g_{ij}\mathcal{L}^{\alpha}.\label{1.23}
\end{eqnarray}
Notice that $T_{0i} \ne T_{i0}$. Using the equations of motion (\ref{1.18}), (\ref{1.19}) one can  explicitly  verify the conservation law,
\begin{equation}
\partial^{\mu}T_{\mu\nu}= 0. \label{1.24.1}
\end{equation}
Hence $T_{\mu\nu}$ is a conserved but non-symmetric energy-momentum tensor.

    Once we have the forms of $T_{00}$ and $T_{0i}$ we can easily obtain the expression for the momenta $P_{\mu}$ and the angular momenta $M_{\mu\nu}$. They are related to the components of the energy-momentum tensor as
\begin{eqnarray}
P_{\mu}&=&\int d^3 x \ T_{0\mu},\label{1.24.2}\\
M_{\mu\nu}&=&\int d^3 x \  (T_{0\mu}x_{\nu} - T_{0\nu}x_{\mu}).
\label{1.25}
\end{eqnarray}
By using (\ref{1.20}) and (\ref{1.21}) we get,
\begin{eqnarray}
P_{0}&=& \int d^3 x \ \sqrt{(\rho^2+ \frac{a^2}{\rho^{\alpha-1}})}\sqrt{1 + (\partial_i \theta)^2}, \label{1.26}\\
P_{i}&=& \int d^3 x \  \theta \ \partial_{i} \rho,\label{1.27} \\
M_{0i}&=& \int d^3 x \  \sqrt{(\rho^2 + \frac{a^2}{\rho^{\alpha-1}})}\sqrt{1 + (\partial_i \theta)^2} \  x_{i} - \theta \ \partial_{i} \rho \ x_{0},\label{1.28}\\
M_{ij}&=& \int d^3 x \  (\theta \ \partial_{i}\rho \  x_{j} -
\theta \  \partial_{j} \rho \ x_{i}). \label{1.29}
\end{eqnarray}
 Using the Poisson bracket (\ref{0.9.1}) we are able to compute the following algebra, 
\begin{eqnarray}
\{M_{ij},M_{kl}\}&=&  (g_{jk}M_{il} - g_{ik}M_{jl} - g_{il}M_{kj} + g_{jl}M_{ki}),\label{1.30}\\
\{M_{oi},M_{kl}\}&=& (g_{ik}M_{0l} - g_{il}M_{0k}),\label{1.31}\\
\{M_{0i},M_{oj}\}&=& - g_{00}\int d^3x \ (\theta \
\partial_{i}\rho \  x_{j} - \theta \  \partial_{j} \rho \ x_{i})
\left(1 -
\alpha\left(\frac{1-\alpha}{2}\right)\frac{a^2}{\rho^{\alpha+1}}\right).\label{1.32}
\end{eqnarray}
Similarlly the algebra between $P_{\mu}$-$M_{\mu\nu}$ is given by
\begin{eqnarray}
\{M_{0i},P_{j}\}&=& P_{0}g_{ij},\label{1.33}\\
\{M_{ij},P_{k}\}&=& g_{jk}P_{i} - g_{ik}P_{j},\label{1.34}\\
\{M_{0i},P_{0}\}&=& -g_{00} \int d^3 x \  \theta \ \partial_{i}
\rho \ \left(1-
\alpha\left(\frac{1-\alpha}{2}\right)\frac{a^2}{\rho^{\alpha+1}}\right).\label{1.35}
\end{eqnarray}
Finally, the algebra between $P_{\mu}$-$P_{\nu}$ is found out to be,
\begin{eqnarray}
\{P_{\mu},P_{\nu}\}&=& 0.\label{1.36}
\end{eqnarray}
Concentrate on the two Poisson brackets (\ref{1.32}) and (\ref{1.35}). We find that for $\alpha =1$ the complete Poincare algebra is satisfied. This corresponds to the usual Chaplygin Poincare algebra \cite{bg}. However for $\alpha\ne 1$
(which corresponds to the GCG model) the Poincare algebra closes only in the large density limit $(\rho>>1)$. It is, however, reassuring to note that the Schwinger condition,      
\begin{equation}
\{T_{00}({\bf x}),T_{00}({\bf y})\} = (T_{i0}({\bf x}) +
T_{i0}({\bf y}))\partial^{x}_{i}\delta({\bf x}-{\bf
y}),\label{1.37.1}
\end{equation}
is satisfied for any $\alpha$ and $\rho$.
\section{Conclusions:}
To conclude, we have studied various aspects of the Generalized Chaplygin Gas (GCG) models. In the nonrelativistic regime, we have constructed a general form
of the Lagrangian for GCG, that obeys the generalized equation of state. Different parameterizations of this master Lagrangian yield different {\it{inequivalent}} models for GCG, such as the one studied here and the one in \cite{gcg}.
In this sense the construction of nonrelativistic GCG is not unique.
Naturally, the same conclusion extends for a relativistic formulation 
 of GCG. 

  For the relativistic scenario, we have proposed a Born-Infeld like
 model for GCG, which in the nonrelativistic limit, reduces to the conventional GCG. However, unlike the usual $\alpha=1$ Chaplygin gas case, the construction
 of a relativistic GCG model, including both density field and velocity potential is nontrivial. In this context, our model reduces to the usual one, quoted
 in literature \cite{jac} for $\alpha =1$ and also has the correct nonrelativistic limit. However the Poincare algebra closes only in the limit of large 
density. 

 The relativistic Lagrangian formulation for GCG, initiated here, opens up
 a host of avenues for future study. First, we wish to extend our previous 
 study\cite{bg} of the Chaplygin matter in FRW spacetime, to the case of a
 GCG, to observe its cosmological implications. Next, our aim is to analyse
 symmetry properties of GCG (first elucidated in \cite{jac} for usual Chaplygin gas), as well as its connection to D-branes.       
 

 \end{document}